\documentclass[12pt,preprint]{aastex} 

\def\tighttable{\def\baselinestretch{1.0}}
\def\ergsec{ergs s$^{-1}$}
\def\Msun{M$_{\odot}$}
\def\HST{{\it HST}}

\def\H{{$H_{160}$}}

\def\cl1056{CL1056$-$03}
\newcommand {\lya}{Ly$\alpha$}

\shorttitle{High-redshift galaxies}
\shortauthors{Zheng et al.}
\slugcomment{Submitted October 13, 2008\\Accepted for publication in {\it The Astrophysical Journal}}
\begin{document}
\title{Bright Strongly Lensed Galaxies at Redshift $\bf z 
\sim 6-7$ behind the Clusters Abell 1703 and CL0024+16\altaffilmark{1}} 
\author{
W. Zheng\altaffilmark{2}, 
L. D. Bradley\altaffilmark{2},
R. J. Bouwens\altaffilmark{3},
H. C. Ford\altaffilmark{2},
G. D. Illingworth\altaffilmark{3},
N. Ben\'{\i}tez\altaffilmark{4},
T. Broadhurst\altaffilmark{5},
B. Frye\altaffilmark{6},
L. Infante\altaffilmark{7},
M. J. Jee\altaffilmark{8},
V. Motta\altaffilmark{9}, 
X.W. Shu\altaffilmark{2,10}, 
and
A. Zitrin\altaffilmark{5}
} 
\altaffiltext{1}{Based on observations with the NASA/ESA Hubble Space 
Telescope, obtained at the
Space Telescope Science Institute, which is operated by the Association of 
Universities of Research in Astronomy, Inc., under NASA contract NAS5-26555, 
and at the Gemini Observatory, which is operated by the
Association of Universities for Research in Astronomy, Inc., under a cooperative agreement
with the NSF on behalf of the Gemini partnership: the National Science Foundation (United
States), the Particle Physics and Astronomy Research Council (United Kingdom), the
National Research Council (Canada), CONICYT (Chile), the Australian Research Council
(Australia), CNPq (Brazil) and CONICET (Argentina).
}
\altaffiltext{2}{Department of Physics and
Astronomy, The Johns Hopkins University, Baltimore, MD 21218}
\altaffiltext{3}{Lick Observatory, University of California, Santa Cruz, 
CA 95064}
\altaffiltext{4}{Instituto de Matem\'{a}ticas y F\'{\i}sica
Fundamental (CSIC), C/Serrano 113-bis, 28006, Madrid, Spain}
\altaffiltext{5}{School of Physics and Astronomy, University of Tel Aviv, 69978, Israel}
\altaffiltext{6}{School of Physical Sciences, Dublin City University, Dublin 9, Ireland}
\altaffiltext{7}{Departmento de Astronom\'{\i}a y Astrof\'{\i}sica,
Pontificia Universidad Cat\'olica de Chile, Santiago 22, Chile}
\altaffiltext{8}{Department of Physics, University of California, Davis, CA 95616}
\altaffiltext{9}{Departamento de F\'{\i}sica y Astronom\'{\i}a, Universidad de Valpara\'{\i}so, 
Valpara\'{\i}so, Chile}
\altaffiltext{10}{Center for Astrophysics,
University of Science and Technology of China, Hefei, Anhui 230026, China}

\begin{abstract} 
We report on the discovery of three bright, strongly-lensed 
objects behind Abell 1703 and CL0024+16 from a dropout search over 25 square 
arcminutes of deep NICMOS data, with deep ACS optical coverage.  They are
undetected in the deep ACS images below 8500 \AA\ and have clear detections 
in the $J$ and $H$ bands. Fits to the ACS, NICMOS and IRAC data yield robust
photometric redshifts in the range $z \sim 6 - 7$ and largely rule out the 
possibility that they are low-redshift interlopers. All three objects are 
extended, and resolved into a pair of bright knots. The bright
$i$-band dropout in Abell 1703 has an $H$-band AB magnitude of 23.9, which 
makes it one of the brightest known galaxy candidates at $z > 5.5$.   
Our model fits suggest a young, massive galaxy 
only $\sim 60$ million years old with a mass of $\sim 10^{10} M_{\sun}$.
The dropout galaxy candidates behind CL0024+16 are separated by 2.5\arcsec\ 
($\sim 2$ kpc in the source plane), and have $H$-band AB magnitudes of 25.0 
and 25.6. Lensing models of CL0024+16 suggest that the objects have 
comparable intrinsic magnitudes of AB $\sim 27.3$, approximately one 
magnitude fainter than L* at $z \sim 6.5$.  Their similar redshifts, 
spectral energy distribution, and luminosities, coupled with their very 
close proximity on the sky, suggest that they are 
spatially associated, and plausibly are 
physically bound. Combining this sample with two previously reported,
similarly magnified galaxy candidates at $z \sim 6-8$, 
we find that complex systems with dual nuclei may be a common feature of 
high-redshift galaxies.
\end{abstract}

\section{INTRODUCTION}
 
Deep observations and surveys with the Hubble Space Telescope (\HST),
such as GOODS \citep{G04,goods}, the HUDF \citep{hudf}, the HUDF05
\citep{oesch}, and the HUDF parallel observations \citep{bouwensudfp}
have greatly enriched our knowledge of the distant Universe, enabling
the discovery of more than 600 $i_{775}-$band dropouts
\citep[$i$-dropout;][]{bouwens,bouwens3}, and eight $z_{850}-$band
dropouts \citep[$z$-dropout;][]{B08} which are believed to be galaxies
at $z\sim 6$ and 7, respectively.  An L* galaxy of $M=-20.3$ at $z
\sim 6$ exhibits an apparent magnitude $\sim 26.8$ (AB, and hereafter); 
consequently, the majority of $z>5.5$ galaxies are fainter than AB=26. 

Bright high-redshift galaxies are rare.  Ten $i$-dropouts brighter
than AB $= 25.5$ were found in $\sim 400$ square arcminutes of GOODS,
HUDF, and HUDF parallel fields \citep{bouwens,bouwens3}. Two of them
have been confirmed to be high-redshift objects spectroscopically
\citep{bunker,stanway1}.  In a survey of the Subaru Deep Field 
of 767 square arcminutes \citep{subaru,taniguchi}, 18 $i$-dropout
galaxies were spectroscopically confirmed to be at $z \sim 5.7$, and
nine at $z \sim 6.6$, but the brightest among them is only at
magnitude of 25.4. In a recent survey that covers one square degree in
the field of the Subaru/XMM-Newton Deep Survey \citep{ouchi},
17 \lya\ emitters (LAEs) at $z\sim 5.7$ were found, of which only one
is brighter than AB=25.5 in the $z$ band. At least
one bright LAE has been found at $z \sim 6.6$ \citep[$z$-band magnitude 25.4,
narrow band (921nm)  magnitude 23.5;][]{himiko}.  Combining the results of 
these three surveys, the brightest $z \sim 6$ candidate over 5000 square
arcminutes is at AB=24.5 (in broad bands -- LAEs are intrinsically brighter
in a narrow band). 

Gravitational lensing by massive galaxy clusters enables 
the detection of high-redshift galaxies that are fainter than can be found 
even in very deep surveys. In addition, lensed galaxies can be resolved 
spatially and, if bright enough, can be candidates for spectroscopy
\citep{hu,kneib,pello,schaerer,richard06,frye,zd1,bouwens4,richard}. 
The number of $z \gtrsim 6$ candidates reported, however, has varied
considerably depending on how aggressive the selection criteria are
\citep[and consequently how reliable the detected sources are: see,
  $e.g.$, discussions in][]{bouwens4}.  The \HST\ images of clusters
are the best in terms of spatial resolution and photometric accuracy,
thus permitting the most reliable selections of high-redshift
galaxies.  One salient example is the three $i$-dropout arclets in
Abell 2218 found by \citet{kneib}.  These arclets represent different
views (multiple images) of the same source at $z=6.75 \pm
0.1$. \citet{bouwens1} found an $i$-dropout object in the field of
cluster RDS1252$-$29 as bright as AB=24.2, and follow-up spectroscopy
\citep{dow} confirmed it as a Lyman-break galaxy at $z=5.515$ and
revealed stellar absorption lines in the continuum.  Recently,
\citet{zd1} discovered a galaxy candidate at $z \sim 7.6$ in the field
of Abell 1689 that is spatially resolved and has an observed $H$-band
magnitude of 24.7. The cluster lensing magnifies this galaxy (A1689-zD1)
by a factor of $\sim 10$.

Taking advantage of the NICMOS data of 21 arcmin$^2$ over 11 galaxy
clusters (including Abell 1689), \citet{bouwens4} searched for
strongly lensed $z$-dropout galaxies at $z \gtrsim 7$ and found that A1689-zD1 
is the only robust candidate in these fields. This suggests that bright
high-redshift galaxies may be rare in cluster fields, even given the smaller 
field size of cluster regions. Since the amount
of NICMOS data available over cluster fields is still very limited,
there is reason to look at the surface density of such objects over
larger areas and at somewhat lower redshifts, in order to overcome the
effect of small number statistics.

The purpose of this paper is to report on the results of a search for
bright ($H$-band magnitude $\leq25.6$), highly magnified $i$- and $z$-dropout 
galaxies over a larger area than was used by \citet{bouwens4}.
Such 
sources provide us with a unique opportunity to examine the properties
of lower-luminosity galaxies in detail.  The lower-luminosity galaxy
population is of great interest because it appears 
to dominate the UV luminosity density, SFR density, and production of ionizing
($\lambda_0\lesssim912$\AA) photons
\citep{yan4,yoshida,hudf,bouwens,bouwens3,oesch,reddy,frye,chen}. 
In particular, we consider $\sim 4$ square arcminutes of additional search 
area over CL0024+16 and CL1056$-$03 obtained in a recent NICMOS program
HST-GO10874 (PI W. Zheng).  These areas were not included in the $z$-dropout 
search of \citet{bouwens4}. 

The plan for this paper is as follows.  In \S2, we describe the
observational data.  In \S3, we outline our search criteria and
summarize the results of our search for bright $z \sim 6-7$ galaxies.
In \S4, we use stellar-population modeling and the available lensing
models to remark on the physical properties of the high-redshift
galaxy candidates we are examining, and in \S5, we summarize our results.

\section{OBSERVATIONS} 

A summary of the \HST\ ACS, WFPC2, and NICMOS imaging observations
used here to search for $z \sim 6-7$ $i$- or $z$-dropout galaxies is
given in Table~\ref{tbl-1}.  Of particular interest are new NICMOS
observations acquired over CL0024+16 (two NICMOS NIC3 pointings) and
CL1056$-$03 (three NICMOS NIC3 pointings) from the program
HST-GO10874. These observations were taken to better quantify the
volume density of faint galaxies at $z \sim 7$.  Each NICMOS image is
only 52\arcsec\ $\times$ 52\arcsec\ in size and covers just a portion
of the total area with deep ACS data over these two clusters. 

The NICMOS/NIC3 images of CL0024+16 were obtained in 2007 July,
with exposures of 9.4 ksec in both the F110W and F160W band.  The
NICMOS/NIC3 images of CL1056$-$03 were obtained in 2007 January and December, 
and 2008 January, with exposures of 6.6-8 ksec in
both the F110W and F160W band.  We used a custom pipeline written at
the University of California Santa Cruz to process the NICMOS images
\citep[``nicred'';][]{magee} into a final mosaic.  The $5 \sigma$
limiting magnitudes for these data are approximately 26.7.   

CL0024+16 was observed in 2004 November in six broad optical bands:
F435W, F475W, F555W, F625W, F775W and F850LP, including 16.3 ksec in
the $z$-band and 10.1 ksec in the $i$-band.  CL1056$-$03 was observed in
2004 March in the F606W, F775W and F850LP bands, including 17.7 ksec in the 
$z$-band and 17.3 ksec in the $i$-band. All the ACS science images over 
these fields were processed with our pipeline APSIS \citep{apsis} and 
registered with the available NICMOS data. 

We retrieved archival Spitzer/IRAC images of the field of
Abell 1703 and CL0024+16, observed in 2007 January and 2003 December, 
respectively. The mosaic images are in the form of
Post-BCD (Basic Calibrated Data), produced by the Spitzer pipeline 
calibration MOPEX. For Abell 1703, the exposure times were 6.2 ksec
for both the 3.6 and 4.5 $\mu$m bands, 
reaching a $5 \sigma$ limiting magnitude of $\sim 24.7$. 
For CL0024+16, the exposure times were 2.7 ksec in the 3.6$\mu$m band and 4 ksec in the 4.5 $\mu$m band, reaching a
$5 \sigma$ limiting magnitude of $\sim 24.5$.  

We also searched for bright $z\sim6$ $i$-dropouts in the cluster data
considered by \citet{bouwens4}, who had used these data to search for
$z$- and $J$-band dropouts at $z\geq7$ and include NICMOS observations over
11 high-redshift clusters (Abell 1689, Abell 1703, Abell 1835, Abell
2218, Abell 2219, Abell 2390, Abell 2667, AC114, CL0024+16, 1E0657-56,
and MS1358+61).  Most of these clusters have deep ACS and WFPC2
optical data to ensure that the candidates we select at $z \sim 6$ as
$i$-dropouts are not detected in the optical to very faint
limits.  We refer to \citet{bouwens4} for more details on the
reductions and processing of those data.  Since we will be reporting
on a bright $z \sim 6$ candidate behind Abell 1703, we will summarize
its ACS + NICMOS observations as well. NICMOS/NIC3 images of Abell 1703 
were obtained in 2007 July, with exposures of 2.6 ksec in both the F110W
and F160W band.  The cluster was observed at three other positions in
the F110W band (see Fig. 1).

Independently as part of another program to search for high-redshift
galaxies, we have acquired deep IR images of these cluster fields
using a wide variety of different ground-based telescopes
(Gemini-North/NIRI, Magellan/PANIC, and the VLT/ISAAC).  Abell 1703
was observed with the Gemini-N/NIRI on 2003 June 11, with a total
exposure time of 7.6 ksec in the $J$ band, and a $5 \sigma$ limiting 
magnitude of 23.3. The Magellan/PANIC observations of CL0024+16 were 
made in 2003
October 3 and 5: the $J$-band observations took 17.8 ksec, and the
image reaches a $5\sigma$ limit of AB$\sim 26$.  The infrared images
were processed with the IRAF/XDIMSUM package. Images taken on each
night and with the same instrument configuration were sky-subtracted,
shifted by the amount determined using the positions of a bright point
source in dithered images, and combined.  We select dropout galaxies
when they are not detected beyond $2 \sigma$ level, and the color
decrements are more than two magnitudes. While these ground-based 
observations are shallower than the NICMOS data, they
were used to tentatively identify the $i$-dropout galaxies.  In a
forthcoming paper \citep{zheng} we will be presenting a few of these
$z$-dropout galaxies from these wider-area ground-based data. 

\section{SELECTION OF BRIGHT $i$- AND $z$-BAND DROPOUTS} 

Source catalogs were produced using SExtractor \citep{sex}.  We
coadded the available NICMOS images for the clusters under study to
create the detection image \citep{szalay}.  We then measured the
fluxes on each of the ACS, WFPC2, and NICMOS images using the same
scalable apertures, after PSF-correcting the higher resolution ACS and
WFPC2 images to match the NICMOS images.  As in \citet{bouwens4}, we
measure colors from smaller scalable apertures \citep[Kron factor of
1.2; ][]{kron}, typically $\sim 0.3$\arcsec.  We then correct the fluxes 
measured in
these smaller apertures to total magnitudes by comparing the light
inside a larger scalable aperture (Kron factor of 2.5) of $\sim 0.6$\arcsec, 
on the detection image to that inside a smaller aperture.

Our color selection is based on the magnitudes measured in
small-scalable apertures. The selection criteria for $i$-dropouts are similar to that of \citet{G04}
and \citet{bouwens}, 
but with some variations: \newline
($i_{775}-z_{850}) > (1.3 + 0.9 (z_{850} - J_{110}))$ \newline
or 
\newline
($i_{775} - z_{850}) > 2.0 \wedge (z_{850} - J_{110}) < 0.8$
\newline
where $\wedge$ represents the logical {\bf AND} symbol.  In addition,
we require that objects are not detected ($< 2\sigma$) in all bluer
bands to exclude lower-redshift interlopers. Also, selected dropout galaxies
must have a star/galaxy index $< 0.8$ and an image dimension greater
than three pixels to exclude stars and artifacts.

The selection criterion for the $z$-dropouts \citep{B08} is:
\newline
$ (z_{850}  - J_{110}) > 0.8) \wedge $ \newline
$(z_{850} - J_{110}) > (0.8 + 0.4 (J_{110} -  H_{160})) \wedge $ \newline
$(J_{110} -  H_{160})) < 1.2$ \newline
Also, we require that objects are non-detected in all the bands shortward 
of 8500\AA. 

In cases of non-detection in a shorter-wavelength band, we define the color
decrement by setting the flux in that band to be zero, with an
error that is equal to the $1 \sigma$ detection limit.  
The non-detection in the $r,g,V$ bands is important, as some sources are 
weakly detected in some of
these bands. We found five other objects that were weakly detected in
the shorter-wavelength bands. Their large $i-z$ color decrements are probably
produced by a Balmer decrement instead of a \lya\ edge, and therefore
the objects are at low redshift, {e.g.} $z \sim 1$. Without deep images in 
the shorter-wavelength
bands, these objects may be considered as ``high-redshift
candidates''. This may explain why some other searches yield a
considerable number of dropout galaxies.

Finally, we set an $H$-band magnitude limit of 25.6 to identify only
the brightest sources in our survey.  Applying the above selection
criteria across our search fields, we identify three new dropout
galaxies in total.  One source, called A1703-iD1, is found in
Abell 1703 \citep[see also][who also report on this source]{a1703} and two 
other sources, CL0024-iD1 and CL0024-zD1, are found
over CL0024+16.  We also identified the $z\sim6.5$ $i$-dropout in Abell 2218 
\citep{kneib} and the $z\sim7.6$ $z$-dropout over Abell 1689 \citep{zd1}.  
No bright dropouts are found in \cl1056\ or any other clusters. 
Our three new galaxy candidates are given in Table~\ref{tbl-2}, along with the
previously known sources. The positions of these sources in
Abell 1703 and CL0024+16 are shown with the red circles in
Fig.~\ref{fig1}-\ref{fig2}. Their cutout stamps are
given in Fig.~\ref{fig3}-\ref{fig4}.

For our high-redshift candidates, we used the available Spitzer/IRAC
data to measure fluxes at 3.6 and 4.5 \micron.  Since these
data have a considerably larger PSF than the ACS/NICMOS data, this
poses a challenge for accurate photometry. First, the IRAC PSF ($\sim
1.6$\arcsec\ at 3.6\micron)
is considerably larger than the apertures used in the \HST\ images 
(typically
$\lesssim$ 0.5\arcsec\ in radius).  We used an aperture size of
2.5\arcsec\ diameter, and then applied corrections of 0.56 and 0.6
magnitude for the $3.6 \mu$m and $4.5 \mu$m band, respectively, to
account for flux outside of the aperture.  These corrections were
derived for stellar profiles \cite[see, $e.g.$,][]{zd1}. In the field
of A1703-iD1, there are two nearby sources of comparable brightness
approximately 2\arcsec\ away (Fig. 3). We ran the task {\it Galfit}
\citep{galfit} to deblend the source fluxes with fixed source
positions and a standard PSF profile of FWHM=1.6\arcsec. The derived 
magnitudes are presented in Table 2.

A1703-iD1 is a prominent $i$-dropout object, with a $z$-band magnitude
AB= 24.2.  As shown in Fig. 3, it is well detected in the ACS $z$-band
and NICMOS $J$ and $H$ bands, but not in the $i$-band and the other
bands at shorter wavelengths.  The object is detected in the Gemini-N
image in the $J$ band, with a magnitude of $24.2 \pm 0.4$.  Our
confidence that this source is a galaxy at $z\sim 6$ rather than a
lower-redshift object \citep{stanway2} is largely motivated by the
source's large color decrement of $>$2.6, which is twice the threshold
in a normal selection ($1.3 - 1.5$).  The possibility that this source
is a dusty galaxy at lower redshift is ruled out as the source is
roughly equally bright in the $z$- and $J$-bands, and entirely
undetected in the $r_{606}$- ($r$) and $i$-bands, which is clearly a
sign of a break in the spectral energy distribution (SED) instead of a
gradual decrease toward bluer wavelengths.  The source is resolved
into two components (magnitudes 24.3 and 24.9; Fig. 3) separated by
$\sim$ 0.45\arcsec.  The only other objects with such colors are
T-dwarfs, which can be ruled out by the fact that the source is
extended.  The FWHM for each component is $\sim 0.4$\arcsec, which is
four times the ACS resolution. This source is detected at low significance 
($< 2 \sigma$) in the F435W band (magnitude $27.7 \pm 0.5$), but not 
in an overlapping band
F475W ($> 28.5$) and F555W and F625W. This weak detection is probably
just an upward statistical fluctuation and does not affect the SED fitting 
results  noticeably.  

The two sources in CL0024 pose an interesting question. These
sources, CL0024-zD1  and  CL0024-iD1, are separated by only 2.5\arcsec.  
As we shall see below they have very similar redshifts and SEDs, and this
small separation corresponds to just 2 kpc in the source plane. Thus
they are likely to be spatially associated.  This will be discussed
further below, but for now we discuss their individual properties.

The source CL0024-iD1 is resolved in both the $J$ and $H$ band images.
Its NICMOS $J$-band magnitude is 25.1 (Table 2). An independent check comes 
from our Magellan/PANIC image data which gives $25.8 \pm 1.1$, which is
consistent with the NICMOS measurement. 
The source CL0024-zD1 appears
to be marginally resolved into two components in the $J_{110}$ band image
(Fig. 4), but not in the $H_{160}$ band.
In the Magellan images, its magnitude is $26.3 \pm 1.3$, which
again is consistent with the NICMOS data. 
The detection in the 4.5$\mu$m band is marginal, and appears to be 
consistent with background fluctuations. 

It is interesting to note that A1689-zD1 \citep[$z\sim 7.6$;][]{zd1}
also consists of two components separated by $ \sim 1$ kpc in the source 
plane. The apparent frequency of
high-redshift galaxies with two bright components is perhaps not
surprising, in that we expect to see bright star-bursting galaxies
imbedded in massive halos that are merging at this very early epoch.

Because of the long time baseline of our ACS, NICMOS, IRAC, and
ground-based observations, none of these sources are likely to be
supernova outbursts or transiting objects. In addition, because the 
sources are detected in multiple bands, it is highly unlikely that 
they are spurious detections. 

\section{DISCUSSION} 

Current studies of the UV luminosity function (LF) at $z \sim 2-6$ suggest 
that lower-luminosity galaxies provide the dominant contribution to the 
overall luminosity density in the UV, the stellar mass density, the ionizing
photon budget necessary for reionization, and the build-up of metals
\citep{yan4,yoshida,hudf,bouwens,bouwens3,reddy,chen}.  As such, clearly 
it is important to quantify the properties of this faint population.
Unfortunately, this can be difficult to do in the field, due to the
integration times required and due to crowding issues in deep IRAC
images. This makes the present sample of highly magnified galaxy candidates 
at $z \sim 6-7$ very useful for investigating the properties of 
lower-luminosity galaxies at high redshifts. 

\subsection{Intrinsic Luminosity} 

To determine the intrinsic properties of the high-redshift galaxy candidates
identified in our search, it is necessary to model the effect of
gravitational lensing on these sources.  Fortunately, both
clusters behind which our galaxy candidates lie have been the subject of
substantial modeling efforts. Abell 1703 ($z=0.258$) is known to have
a large number ($i.e.$, 16) of multiply lensed galaxies and has been
the subject of detailed modeling \citep{limousin}. CL0024+16
($z=0.396$) has also been the subject of several detailed efforts to
model the lensing \citep{kneib3,jee,priya,adi}.  We adopt the model of
\citet{adi} as our fiducial lensing model, but compare with other
models to gauge the approximate uncertainties.  As shown in Table 4,
we adopt a magnification at our source position of 3.1 for A1703-iD1, 
6.3 for CL0024-iD1, and
6.2 for CL0024-zD1. For CL0024+16, a wide range of magnification exists
in the literature. The model of \cite{priya} yields a magnification of 2 
and that of
\cite{jee} predicts a magnification of 17-20. The low magnification value
probably is arose because \cite{priya} assume the cluster mass
profile as singular isothermal, and they used five multiple
images of the same source near the center (as well as weak lensing
constraints). The recent work of \cite{adi} takes advantage of
high-quality, multi-band ACS data and identifies 11 sets of multiple images
covering a wide range of background redshift $0.5 < z < 4.5$, so that the
gradient of mass profile slope can be meaningfully constrained and hence
more reliable magnification measurements are obtained. 
While we have adopted perhaps the most reliable magnification
model, the actual magnification factors in current lensing models 
are sensitive to details in the mass distribution of the foreground 
cluster and thus subject to significant uncertainties.

Adopting the photometric redshifts for A1703-iD1, CL0024-iD1, and
CL0024-zD1 that we will determine in the next section ($i.e.$,
$z=6.0$, $z=6.5$, and $z=6.6$, respectively), we estimate that their
amplification factors are approximately 3.1, 6.3, and 6.2. respectively.  This implies intrinsic $J$-band luminosities for these
sources of AB=25.2, 27.1, and 28.0, respectively.  The two fainter
dropout sources in CL0024+16 are estimated to have intrinsic
magnitudes fainter than L* after demagnification and would not likely
have been found in even the NICMOS data to the HUDF
\citep{bouwenszd,thompson}. A1703-iD1, on the other hand, appears to
be quite bright, having an intrinsic magnitude of $\sim 25$.

\subsection{Spectral Energy Distribution and Photometric Redshift} 

We performed fits to the multiband photometry of these three
sources by using the stellar-population models of \citet{bc03}.
For all of the models, we assume a simple stellar population
(single-burst; SSP) with a \citet{imf} initial mass function with mass
cutoffs of 0.1 and 100 $M_{\sun}$.  We explored models with both solar
($Z = 0.02$) and subsolar ($Z = 0.0004 = Z_{\sun} / 50$)
metallicities.  The models use the obscuration law of \citet{Calzetti}
and are corrected for Lyman-series line-blanketing and photoelectric
absorption following the procedure of \citet{madau}. In the stellar-population 
model fits, we constrain the stellar age to be less than the age of the 
universe at the fit redshift ($e.g.$, 0.8 Gyr at $z =6.5$).

The best-fit stellar-population models are shown in
Fig.~\ref{fig5}-\ref{fig7} and the parameters are given in
Table~\ref{tbl-3}.  While the non-detections are shown in these
figures as 2 $\sigma$ upper limits, we used the measured fluxes,
not the limits, to fit the SEDs.  For both solar and subsolar
metallicities, we found acceptable fits for the models. IRAC
photometry places important constraints on the stellar-population
models as it provides the rest-frame optical magnitudes for
high-redshift galaxies \citep{labbe}. Using the ratio of UV-optical
luminosity one can estimate the galaxy age with considerable
accuracy. Considering the uncertainties associated with the mass models 
themselves and in potential reddening factors, the real errors are likely
somewhat higher than those listed in Table 3.

For A1703-iD1 (Fig. 5), the best-fit
redshifts are $z = 5.9 - 6.0$ with an $1\ \sigma$ error of 0.1. The best-fit 
results, with $\chi^2_{\nu} =1.1$, suggest a high-redshift galaxy with a mass 
of $ 0.9 - 1.2 \times 10^{10} M_{\sun}$ and age of $55 - 64$ Myr.
\citet{a1703} measure a
redshift of $z=5.827$ from Keck spectroscopy, which is broadly consistent 
with our fitting results. We also ran the fits to photometry with the redshift
fixed to $z=5.827$, 
and the results are presented in Table 3 and Fig. 5. Using the spectroscopic redshift 
removes one of the degrees of freedom, but the modeling results still do show 
some dependence on the unknown metallicity: The fit with solar 
metallicity produces a satisfactory result, with $\chi^2_{\nu} =1.1$, but that
with a low metallicity does not fit as well, with $\chi^2_{\nu} =1.7$,
suggesting that the metallicity may well be high in this very luminous 
(3 L*) galaxy, or, if the metallicity is low, that the low-metallicity models may not match the SED very well in these high-redshift galaxies. 
The low-redshift solutions are inconsistent ($\chi^2_{\nu} \sim 10$) with the 
observed photometry. 

For CL0024-iD1 (Fig. 6), again we find that the high-redshift
solutions ($\chi^2_{\nu} = 0.5 - 0.7$) provide a better fit to the
data than the low-redshift solutions ($\chi^2_{\nu} = 2.8 - 4.2$).
Of note, the low-redshift SED models are inconsistent with the
observed $r$- and $i$-band 1~$\sigma$ upper limits and the $z$-
and $J_{110}$-band photometry.  The best-fit redshifts are $z =
6.4 - 6.5$ with an $1\ \sigma$ error of 0.1.  The stellar masses
are $1.5 - 2.1 \times 10^{9} M_{\sun}$ with ages of $43 - 72$ Myr.
The reddenings derived from the fits to both the solar and subsolar
models were $A_V = 0.0$.

For CL0024-zD1 (Fig. 7), we find the high-redshift solutions
($\chi^2_{\nu} \sim 0.1$) are only modestly better than the
low-redshift solutions ($\chi^2_{\nu} = 0.6 - 0.9$).  With the
exception of the $z$-band photometry, which has large uncertainty, the
low-redshift SED models can also reproduce the measured fluxes blueward
of the break in the SED.  We still favor the high-redshift
solutions ($z = 6.5 - 6.6 \pm 0.4$) over the low-redshift models ($z =
1.4 \pm 0.6$).  The stellar masses are $\sim 2-2.5 \times 10^{9}
M_{\sun}$ and the ages $\sim 85$ Myr. The solar and subsolar models
have fit reddenings of $A_V = 0.0$ and $0.2$, respectively.

\subsection{Are CL0024-iD1 and Cl0024-zD1 Physically Related?} 

As noted in \S3, the source CL0024-zD1 is located only 2.5\arcsec\ 
from the CL0024-iD1 counterpart. For our fiducial lensing model where the
amplification is 6.2 (Table 4), they would be separated physically by
only 2.2 kpc.  The two sources are of similar luminosity, differing
by only approximately 0.7 magnitude in both the in $J$- and $H$-bands. The
SED fits suggest very similar redshifts, differing only by
$\Delta z \sim 0.1$ , which makes them essentially identical within the
errors (see Table 3).  The formal designation indicates that one is
an $i$-dropout and one a $z$-dropout -- at $z \sim 6.5$ they have redshifts in
the transition range where small changes in flux in the bands can
result in a change in their dropout designation.  
Given the similar redshifts, SEDs, and luminosities, and the quite small  
separation in projection, it is possible that the two sources are simply 
components of the same source with a physical separation of 2 kpc.
We think it is unlikely they are multiple images of the
same source since our lensing models do not predict such a split pair.
Both sources show dual-core morphology (see \S4.6), and their four components
seem well aligned over a scale of 6\arcsec\ (see the $J$-band image in 
Fig. 4).  
We have kept them as separate entities throughout our discussion but
suspect that it is highly likely that they are spatially associated.
Given the small separation and similarities these objects will likely
be components of a group that will ultimately merge.

\subsection{Surface Density of Bright High-Redshift Candidates} 

We compare our sample (Fig. 8) with the distribution of known galaxy
candidates at $z \sim 6$, based on the LF of \citet{bouwens} and scaled to a search area of 25 square
arcminutes. To a limiting magnitude of 25.6, a survey in an unlensed
field is anticipated to find $\sim$1.4 candidates at $z\sim 6$.  We,
however, find 4-5 such bright $z \sim 6-7$ candidates behind lensing
clusters.  This effectively illustrates how lensing from foreground
clusters can help us identify high-redshift galaxies
at relatively bright magnitudes -- which is helpful for numerous
follow-up studies, including spectroscopy. The observed enhancement
is expected because of the steep slope of the LF expected at higher
luminosities ($i.e.$, $-d(\log d\Phi)/d\log L > 1$: $e.g.$, \citet{br95}).
Such steep slopes are seen brighter than L* where there is an exponential cut-off, 
but not fainter than L*.  
For A1703-iD1, its intrinsic $z$-band magnitude is approximately 25 
after correcting for the $3 \times$ model magnification factor (see Table 4).
According to the $z\sim 6$ LF,
one would expect to require 40-250 square arcminutes to find one such bright candidate. 
Since the discovery of A1703-iD1 was made over an area of 
approximately 7 square arcminutes (after correcting for the $3 \times$ dilution
factor), it would seem to be a rather serendipitous discovery. For the four 
other dropout candidates listed in Table 4,  if we assume a nominal 
magnification factor of 9 in the cluster field, their intrinsic magnitudes are
AB $27-28$. Over two square arcminutes in the source plane, the 
$z\sim 6$ LF predicts approximately 2-5 such galaxies. The four such galaxies 
found are therefore consistent, within the uncertainties, with the known 
distribution of high-redshift galaxies.

\subsection{Rest-frame Optical Fluxes of Lower-Luminosity Galaxy Population} 

Perhaps the most interesting property to examine in our small sample
of highly magnified galaxy candidates is their stellar mass.  The stellar
masses tell us about the past star formation rate in a galaxy and
therefore its likely contribution to the reionization of the universe
as well as the production of metals.  The stellar masses estimated in
the previous section are useful for assessing the contribution of
individual galaxies to the reionization of the universe.  However,
what we really want to know about the galaxy population at $z\sim6-7$
is their stellar mass density -- since we can use this to constrain
the SFR history at $z>7$.  It is difficult to determine this directly
from the above measurements since we only have five high-redshift galaxy 
candidates from which
to derive a mass function \citep[and also the masses we derive will be
  sensitive to uncertainties in the model magnification factors: see,
  $e.g.$,][]{B09}.

A better way of estimating the total stellar mass density is to start
with determinations of the UV LF at $z\sim6-7$ and
then use the measurements of UV-to-optical colors of $z \sim 6-7$ galaxy
candidates to
estimate the stellar mass function of the galaxies (given the link
between rest-frame optical colors and our estimates of stellar
mass). Such colors have been available for the more UV luminous
galaxy candidates at $z \sim 6-7$ \citep{eyles,yan,stark}, but until the 
present have been lacking for the less UV luminous galaxy candidates at 
$z \sim 6-7$. This
is where our sample of highly magnified galaxy candidates at $z \sim 6-7$ is
useful since they provide a luminosity-to-mass calibration for the
less luminous galaxies that have previously lacked colors. 

For the sake of uniformity, we will use the $H-3.6\mu$m colors of
our sample of high-redshift galaxy candidates to represent the rest-frame
UV-to-optical colors.  We plot this color distribution for the four
strongly lensed galaxy candidates in our sample with IRAC fluxes in
Fig.~\ref{fig9}.  We compare this distribution with the colors of more
luminous galaxy candidates at $z \sim 6-7$ identified in the HUDF
\citep{yan5,labbe}.  No significant differences are observed between
the $H-3.6\mu$m colors of our sample and the more
luminous galaxy candidates \citet{labbe} examined in the HUDF. A t-test 
yields  a significance of 51\%, suggesting 
no large differences between the ages (and
mass-to-light ratios) of the lower-luminosity galaxy population at
high redshift and that at higher luminosities.  This has obvious
implications for the total stellar mass density at $z\sim 6-7$ and
therefore the total star formation rate 
density at $z\gtrsim 7$ \citep{chary}, 
since it allows us to estimate the 
fraction of the total stellar mass density at $z \sim 6-7$ that come 
from high- and low-luminosity systems.
Our result does not suggest an increase in the number of low-luminosity
galaxies. It is also consistent with the hypothesis that
more massive galaxies are unresolved aggregates of lower-luminosity galaxies.

\subsection{Morphology of High-Redshift Galaxies} 

Strong lensing enables us to resolve otherwise unobservable spatial
structures in high-redshift galaxies, and thereby investigate the
structure of young galaxies and the physical processes that drive both
their growth and vigorous star formation.  The observed and inferred
physical sizes and separations of the resolved components of the three
sources reported here are given in Table~\ref{tbl-4}.  For A1703-iD1,
there are two knots separated by 0.45\arcsec. Each of them can be
fitted with a two-dimensional Gaussian with FWHM of 0.66\arcsec\ and
0.51\arcsec, respectively. After demagnification (Table 4), this pair
of components is separated by just 0.9 kpc, suggesting a merging event
that drives star formation. Both the candidates in CL0024+16, which 
themselves are probably related as discussed in \S4.3, seem to
consist of two components, which are best seen in the $J_{110}$ image.

It is interesting to note that the $z$-dropout in Abell
1689 (A1689-zD1) also has double nuclei, separated by $\sim 1$ kpc (Table 
4), in the source plane. Double nuclei seem to be almost ubiquitous in 
this sample (see Table 4), and even A2218-iD1 \citep{kneib} appears to 
show significant substructure.
One interpretation is that these
galaxies may consist of multiple star-forming regions that are UV
luminous. The separation between the two sub-structures seems to be
sufficiently large (Table 4) that they may be two merging systems.
For A1689-zD1 or
A1703-iD1 the models of \citet{limousin7,limousin} 
do not predict any counter images for these sources, and 
there is a lack of any luminous foreground sources that would be
responsible for this lensing. Similarly as noted in \S3, our 
models for CL0024+16 do not suggest multiple images for the sources.
The complex and multiple nature of all these sources certainly suggests,
as we would expect from numerical models, that interaction, merging and 
clumpy star formation are all aspects of the very dynamical nature
of the galaxy build-up process at these redshifts.

\section{SUMMARY} 

We have found three bright, strongly lensed $i$- and $z$-dropouts in the fields 
of the massive clusters Abell 1703 \citep[see also][]{a1703} and CL0024+16. Their
large color decrements make the selection of these high-redshift galaxy candidates 
robust (with the exception of CL0024-zD1 for which the evidence is 
little weaker), and detailed cluster mass models enable us to derive the 
intrinsic properties of the high-redshift galaxies. 
The similar redshifts, SEDs, and luminosities, and the quite small
separation in projection, of the two sources in CL0024+16 suggest that they
may be spatially associated with a separation of  $\sim 2$ kpc.
Using stellar-population models to fit the
rest-frame UV and optical fluxes, we derive best-fit stellar masses on
the order of $10^{9-10}$~\Msun. Even after correcting for the lensing
magnification (a factor of 3), A1703-iD1 remains intrinsically
luminous. The three galaxy candidates are extended, with intrinsic sizes 
between 0.2 and 2 kpc, and resolved into a pair of nuclei. They, along with
A1689-zD1 \citep{zd1}, which is also double with component separation 
of $\sim 1$ kpc in the source plane, suggest that merger driven growth 
and star formation may be a defining characteristic of high-redshift
galaxies. Bright resolved galaxies such as these will be important
targets for spectroscopy with the largest ground based telescopes and
future large-aperture space telescopes like {\it JWST}.

\acknowledgments 

ACS was developed under NASA contract NAS 5-32865, and this research has been 
supported by NASA grant NAG5-7697. Infrared observations were made using the
Gemini-North Telescope during the quarter GN-2003B-Q-28. 
We acknowledges the support of \HST\ grants AR10310 and GO10874.
Archival \HST\ data were provided by the Multimission Archive at the Space 
Telescope Science Institute (MAST). We thank the anonymous referee for many 
critical comments, which enable us to improve the quality of this paper 
considerably.

\clearpage 
\centerline{\bf Figure Captions}
\bigskip 
\figcaption{Composite color ($r_{625}$, $i_{775}$, $z_{850}$) ACS image 
  of Abell 1703.  The $z=6$
  critical curves (magnification factor $\mu>75$) for the
  lensing model of \citet{limousin} are overplotted.  The positions of the
  shallow (one orbit each) NICMOS/NIC3 F110W and F160W images
  are illustrated with the four blue boxes and an orange box,
  respectively.  The location of A1703-iD1 is marked with a red
  circle.
\label{fig1}} 

\figcaption{Composite color ($r_{625}$, $i_{775}$, $z_{850}$) ACS
  image of CL0024+16.  The $z=6.5$ critical curves (magnification $\mu
  > 200$) for the lensing model of \citet{adi} are overplotted. The
  blue boxes show the regions with deep NICMOS/NIC3 F110W imaging
  observations.  Two of these NIC3 pointings are from \citet{richard}
  and two are from HST program GO-10874.  The positions of the two
  dropout galaxies CL0024-iD1 and CL0024-zD1 are marked with the red
  circles.
\label{fig2}} 

\figcaption{Cutout images of the $i$-dropout galaxy A1703-iD1. Each
  cutout is 15\arcsec on a side.  The source is detected in the ACS z
  band and NICMOS F110W and F160W bands, but not in the five optical
  bands at shorter wavelengths. The $\chi^2$ image formed from the
  F435W, F475W, F555W, F625W, and F775W optical bands \citep{szalay}
  is presented here and labeled ``Det.''  The $J$-band data taken with
  the Gemini North telescope are also shown. This source is clearly
  resolved into two extended nuclei.  With an $H$-band magnitude of
  23.9, it is one of the most luminous galaxy candidates at $z \sim 6$ 
  ever found.
\label{fig3}} 

\figcaption{Cutout images of the $i$-dropout galaxies CL0024-iD1
  (marked with a red circle) and CL0024-zD1 (green circle). Each cutout
  is 15\arcsec on a side. The $\chi^2$ image formed from the F435W,
  F475W, F555W, F625W, and F775W optical bands \citep{szalay} is also
  presented here and labeled ``Det.''  CL0024-iD1 is detected in the
  ACS $z$-band, NICMOS F110W and F160W bands, and the IRAC 3.6$\mu$m
  and $4.5\mu$m bands while CL0024-zD1 is detected in the NICMOS F110W,F160W 
  and the IRAC 3.6$\mu$m band. Neither source is detected in the five ACS 
  optical bands at shorter wavelengths.  The $J$-band data taken with the
  Walter Baade (Magellan) telescope are also shown. In the $J_{110}$ band,
  each source displays a dual-core structure, and the four components are
  well aligned.
\label{fig4}} 

\figcaption{Best-fit stellar-population models for A1703-iD1.
  The stellar masses are intrinsic values, corrected for the 
  magnification factor of 3 as predicted by the lensing model of
  \citet{limousin}. The vertical bars denote the $1 \sigma$ flux
  uncertainties and the horizontal bars represent the filter
  bandwidths. Non-detection in a given band is marked with a $2 \sigma$
  upper limit in flux and a down arrow.
  The deep ACS data eliminate lower-redshift sources. The
  red and blue curves represent the $z\gtrsim5$ fits with solar
  metallicity (Z=0.02) and subsolar (Z$_{\sun}/50$) metallicity,
  respectively, and the cyan curve for the $z = 5.827$ fits with solar
  metallicity. The green and brown curves are for $z \lesssim 4$ fits
  and have high values of $\chi^{2}_{\nu}$. 
\label{fig5}} 

\figcaption{Best-fit stellar-population models for CL0024-iD1.  The
  stellar masses are intrinsic values, corrected for the magnification
  factor of 6.3 as predicted by the lensing model of \citet{adi}.  
  The green and brown curves are for $z \lesssim 4$ fits
  and have high values of $\chi^{2}_{\nu}$.  See the caption of Fig~\ref{fig5}.
\label{fig6}} 

\figcaption{Best-fit stellar-population models for CL0024-zD1.  The
  stellar masses are corrected for the magnification factors of 6.2 as
  predicted by the lensing model of \citet{adi}.  The $z\gtrsim5$ fit
  gives a much lower value of $\chi_{\nu}^2\sim1$ than for the $z\lesssim4$ fit
  where $\chi_{\nu}^2 \sim2.5$.  Also see the captions of
  Fig~\ref{fig5} and Fig~\ref{fig6}.
\label{fig7}} 

\figcaption{$i$-dropout surface density. The hatched area shows the
  number of $i$-dropouts per 20 square arcminutes in the field, as
  determined by \citet{bouwens3}.  The observed brightness for the
  five dropout galaxies in Table 2 are marked with stars, while their
  intrinsic (demagnified) $H$-band magnitudes are marked with the solid
  circles.  The high magnification results in a number of bright
  sources (though they are intrinsically low luminosity, in
  general). As suggested in a number of studies, cluster magnification
  can result in bright $z\sim 6$ galaxies ($H<25$) that are rare in
  the field (see \S4.3).
\label{fig8}} 

\figcaption{The $H-3.6\mu$m color distribution (red histogram) for the
  lower luminosity, highly magnified galaxy candidates at $z \sim 6.5-7.6$ 
  in our
  selection (Table~\ref{tbl-2}).  This distribution should provide an
  approximate measure of the size of the Balmer break and therefore
  provide us with an approximate measure of its age (though this color
  will also depend upon the dust content).  In general, redder
  galaxies would be expected to be older and bluer galaxies younger.
  Also shown is the color distribution for the more luminous galaxy 
  candidates at $z \sim 6.7-7.2$ in the HUDF \citep[black
    histogram:][]{yan5,labbe}.  Our sample of lensed galaxies has
  similar $H-3.6\mu$m colors to those of the more luminous sample.
  This suggests that the lower-luminosity galaxy population may have
  similar mass-to-light ratios as higher-luminosity galaxies (see
  \S4.4).
  \label{fig9}}
\clearpage 

\setcounter{figure}{0}
\begin{figure}
\plotone{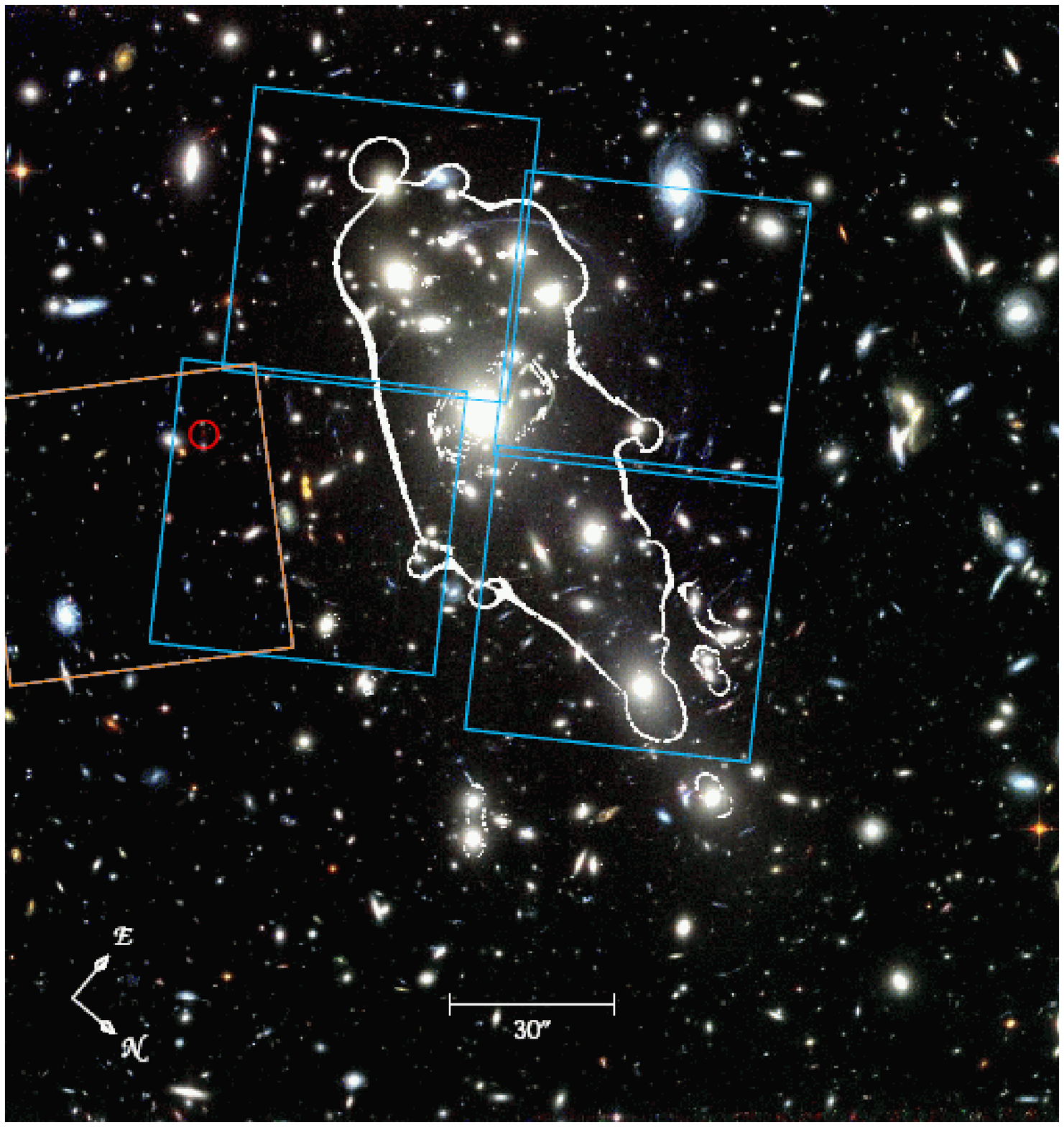}
\end{figure}
\begin{figure}
\plotone{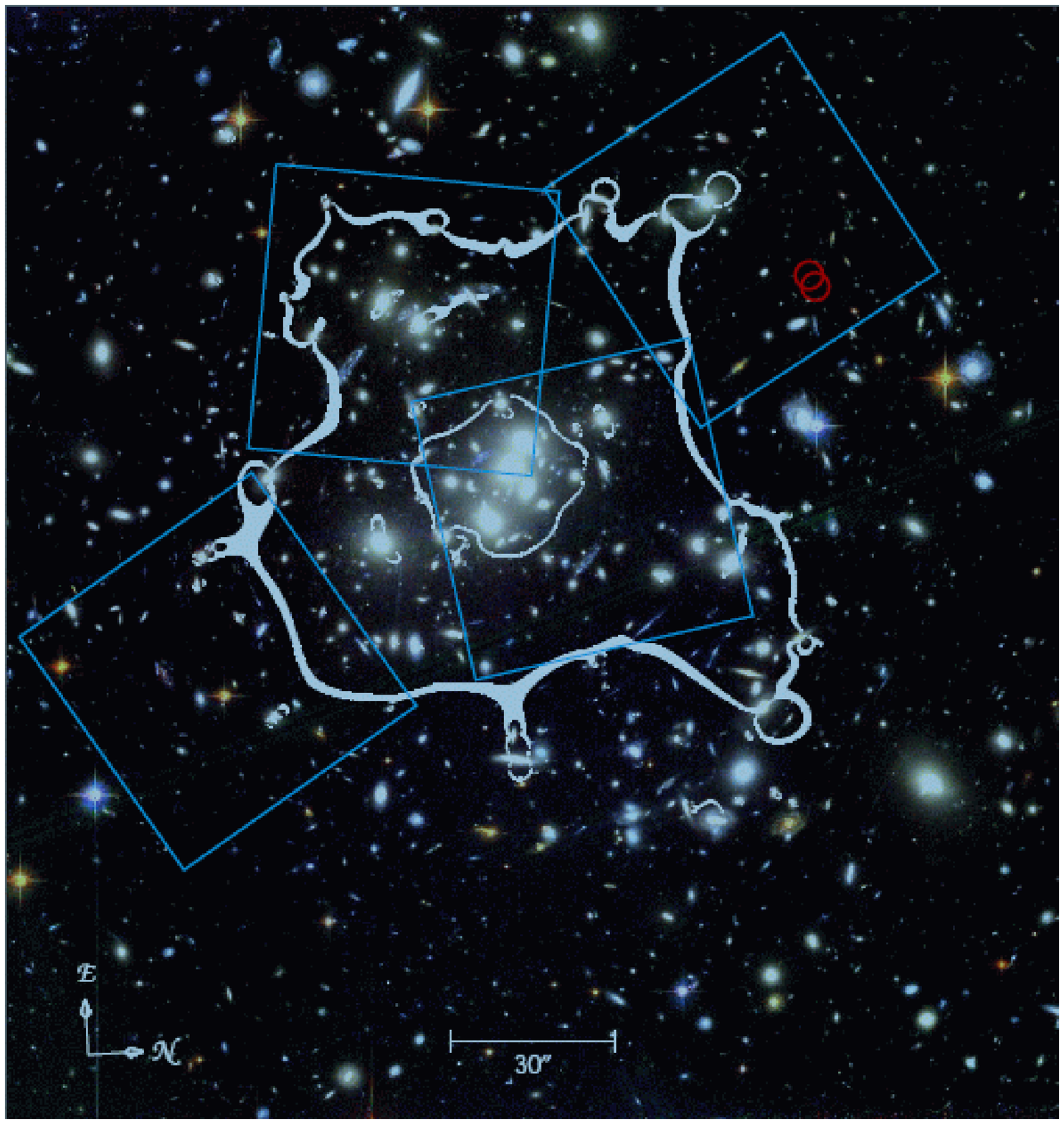}
\end{figure}
\clearpage 
\begin{figure}
\plotone{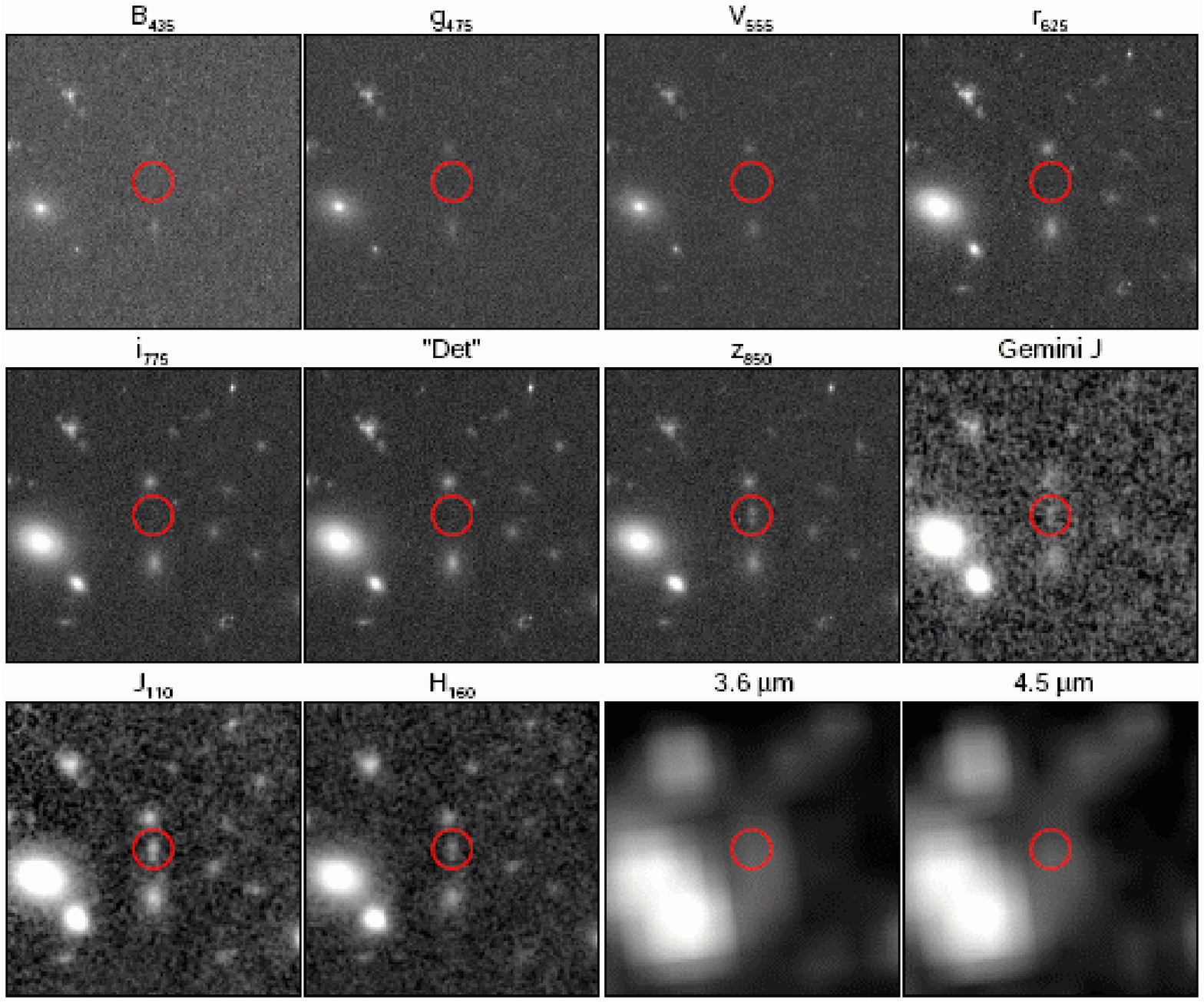}
\end{figure}
\begin{figure}
\plotone{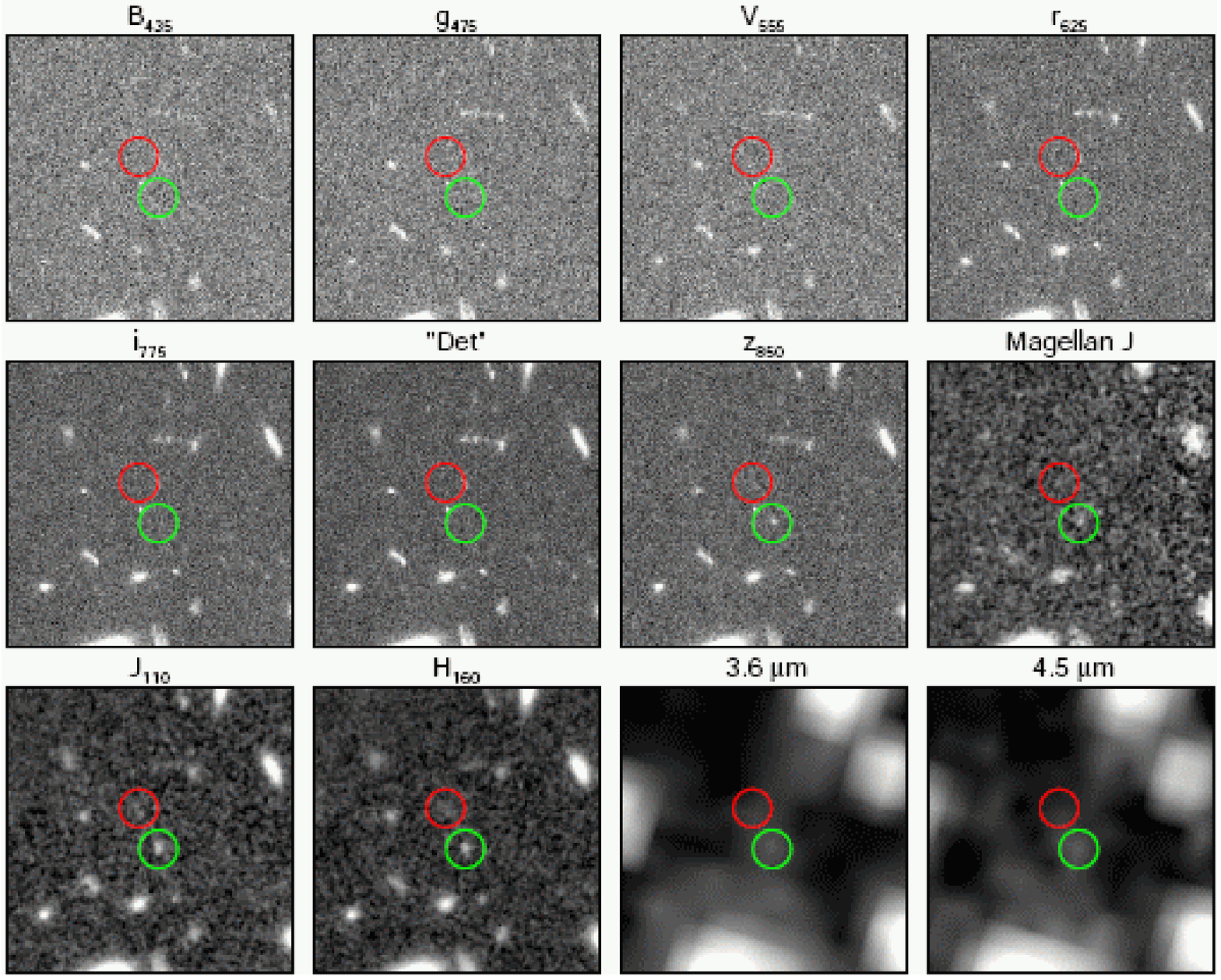}
\end{figure}
\clearpage 
\begin{figure}
\plotone{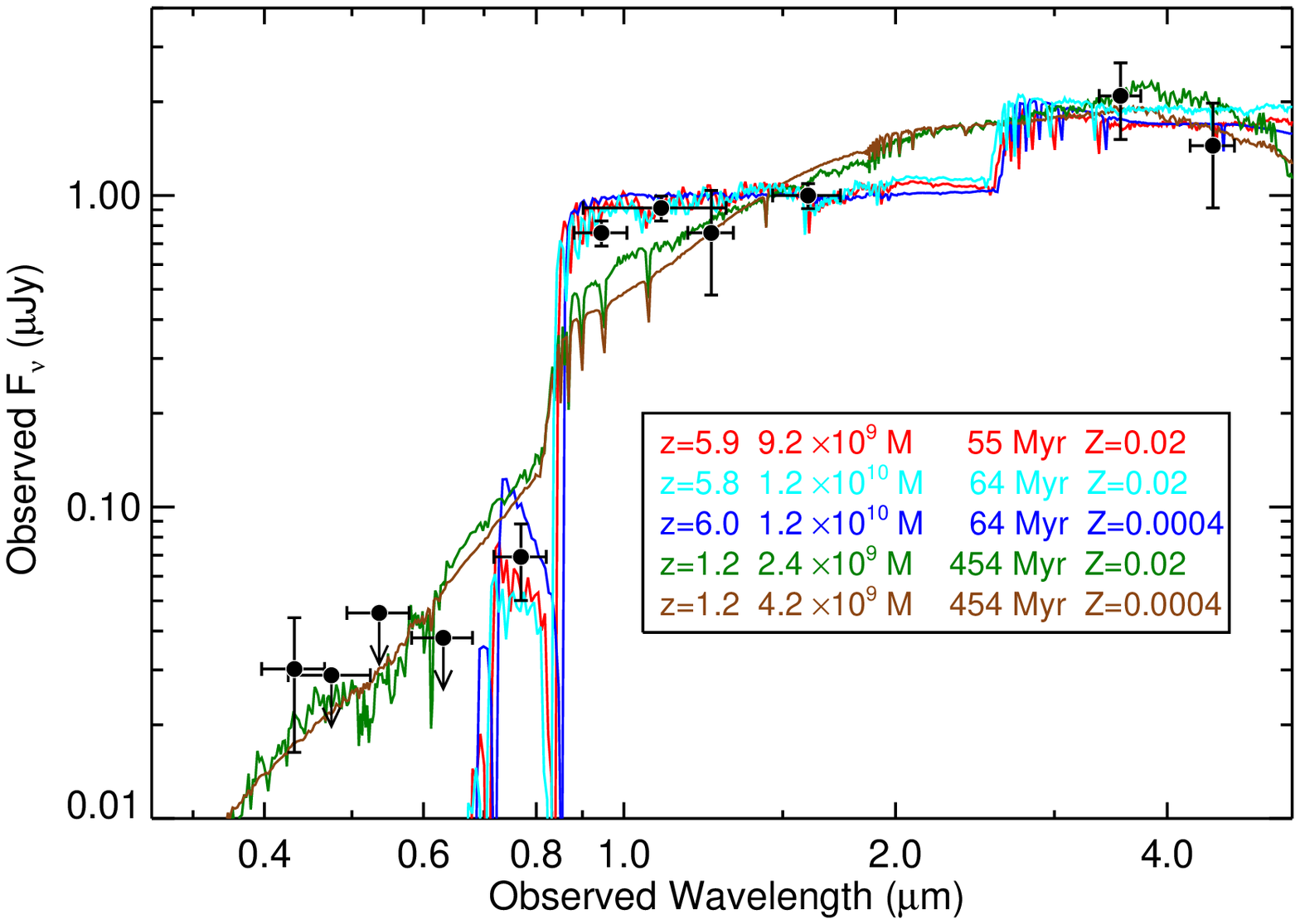}
\end{figure}
\begin{figure}
\plotone{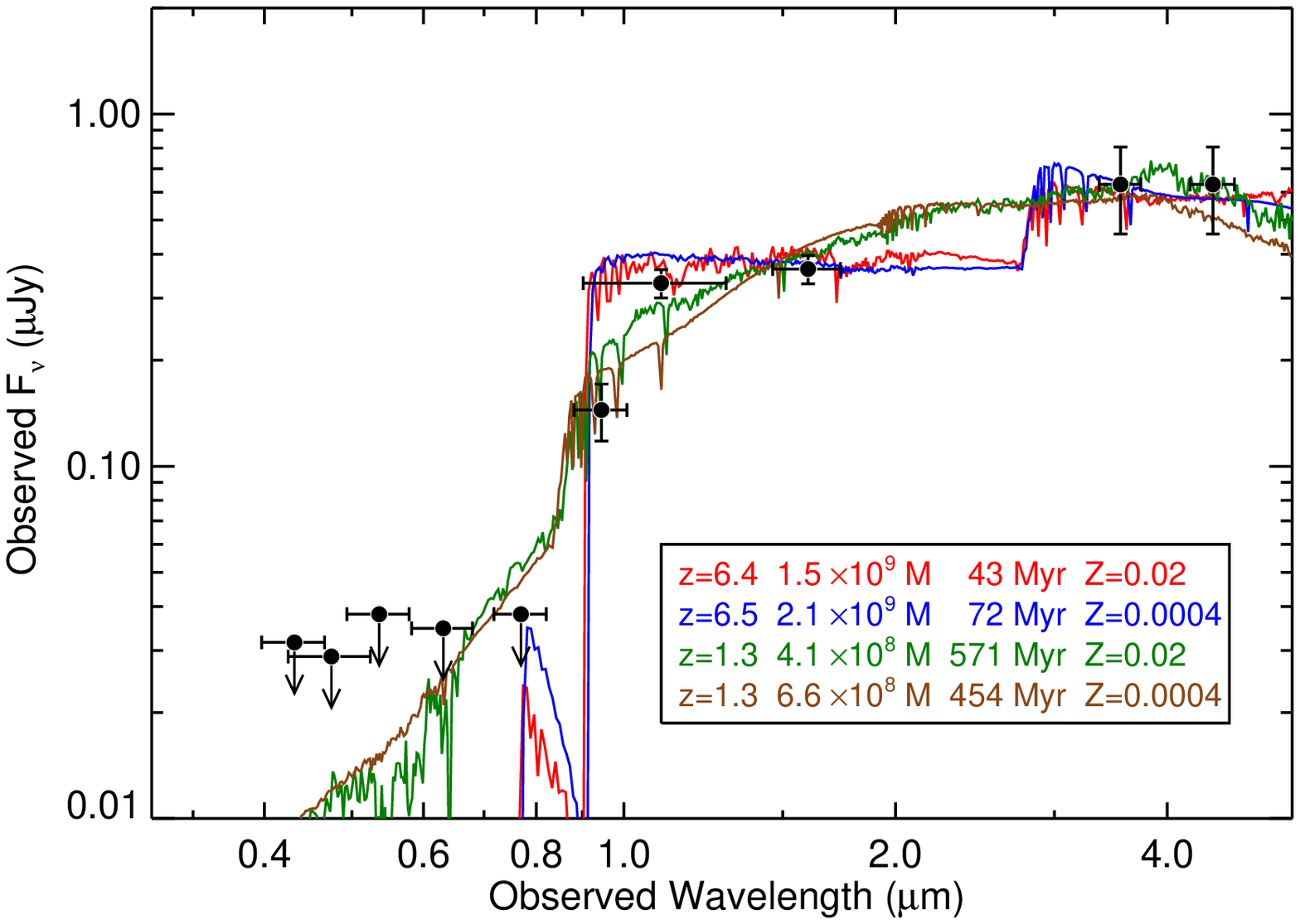}
\end{figure}
\begin{figure}
\plotone{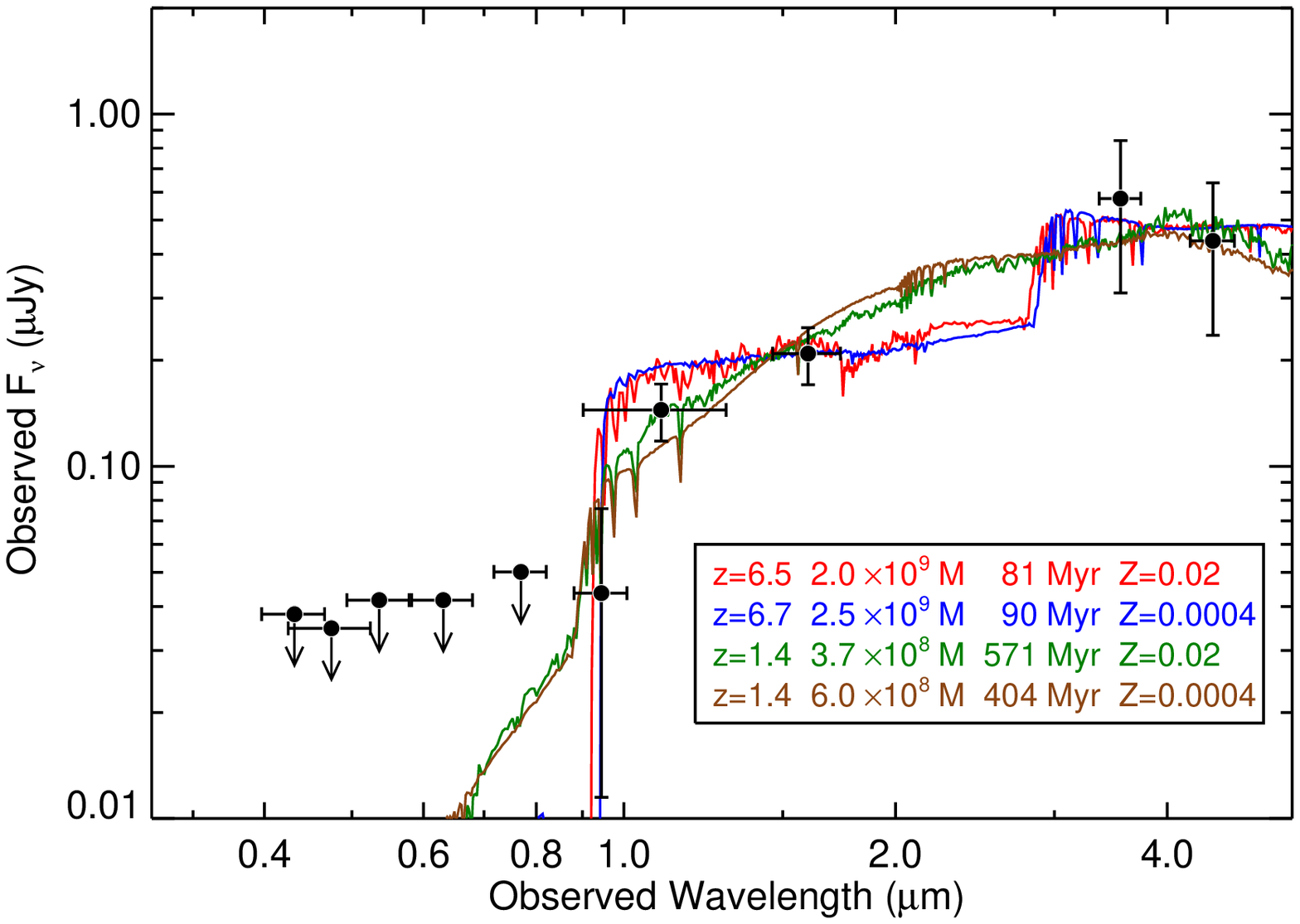}
\end{figure}
\clearpage 
\begin{figure}
\plotone{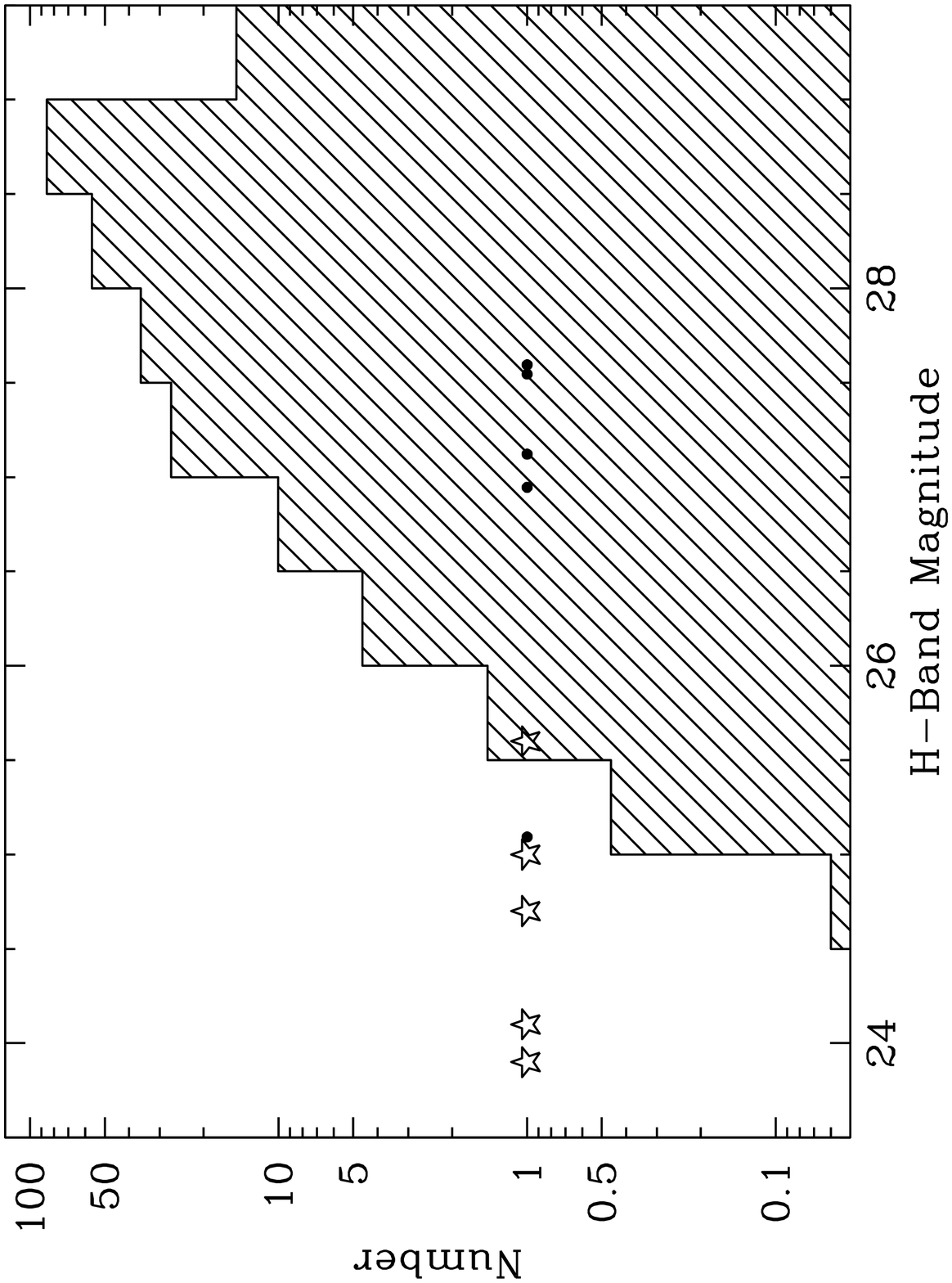}
\end{figure}
\begin{figure}
\plotone{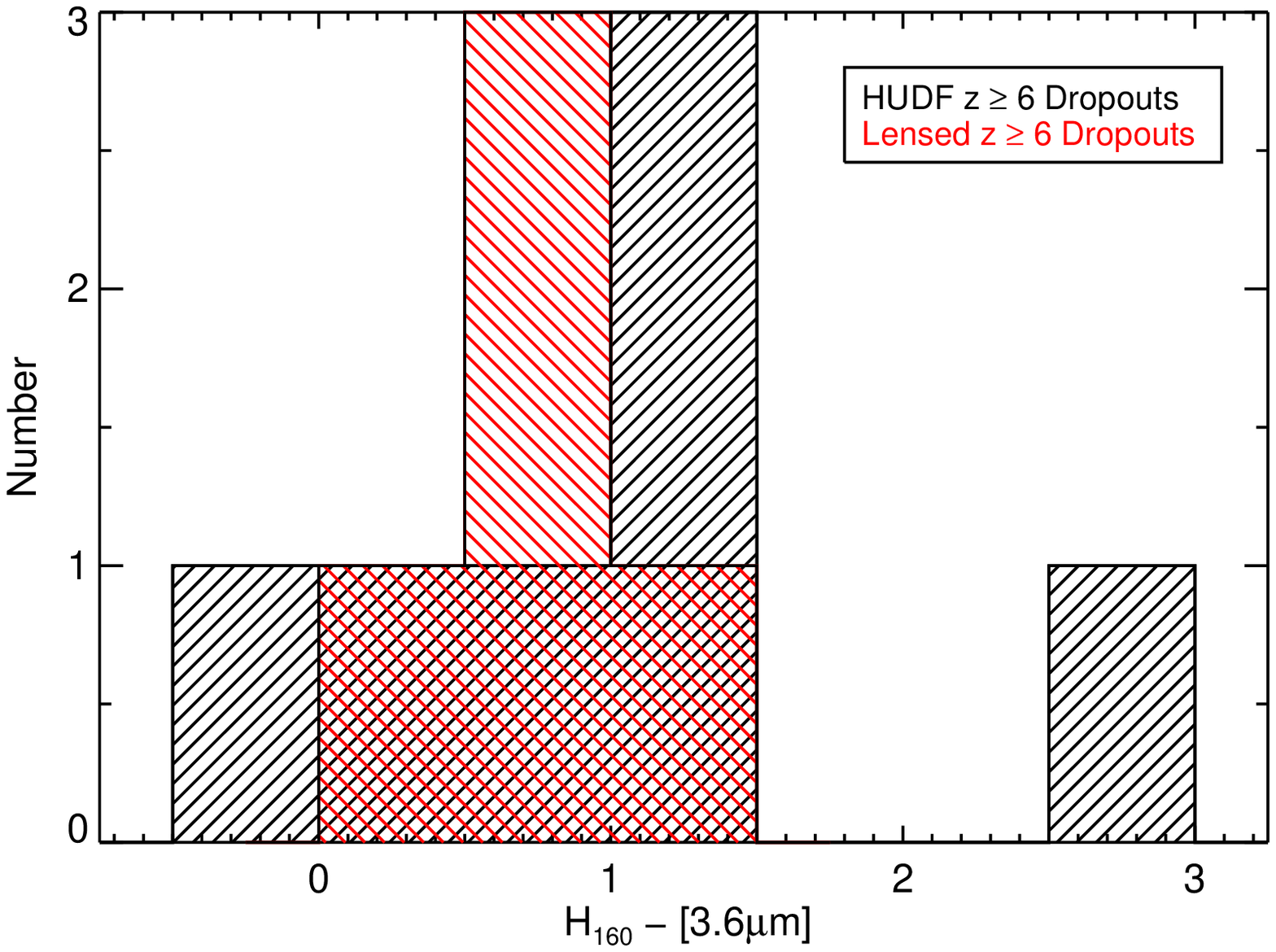}
\end{figure}
{
\begin{deluxetable}{ccccccc} 
\tablecolumns{7}
\tablecaption{\HST\ NICMOS Data around Massive Galaxy Clusters Used to Search 
for $i-$ and $z-$Dropout Galaxies at $z \sim 6-7$\label{tbl-1}}
\tablehead{
\colhead{} & \colhead{Area} & \multicolumn{4}{c}{5$\sigma$ Depth AB Magnitude\tablenotemark{a}} & \colhead{} \\
\colhead{Name} & \colhead{(arcmin$^2$)} & 
\colhead{$i_{775}$\tablenotemark{b}} & \colhead{$z_{850}$} & 
\colhead{$J_{110}$} & \colhead{\H} & \colhead{Ref\tablenotemark{c}}}
\startdata
\multicolumn{7}{c}{New $i$- and $z$-Dropout Search Data (HST GO10874)} \\
CL0024+16 & 1.5 & 27.8 & 27.4 & 26.7 & 26.6 & [1] \\
\cl1056 & 2.2 & 27.5 & 27.3 & 26.6 & 26.5 & [1] \\
\multicolumn{7}{c}{$i$-Dropout Search Data\tablenotemark{d}} \\
MS1358+61 & 1.4 & 28.0 & 27.5 & 26.8 & 26.7 & [2] \\
CL0024+16 & 1.4 & 27.8 & 27.4 & 26.8 & 26.7 & [2] \\
Abell 2218 & 1.2 & 27.9 & 27.6 & 26.8 & 26.7 & [2] \\
Abell 2219 & 1.4 & 27.5 & 27.2 & 26.8 & 26.7 & [2] \\
Abell 2390 & 1.5 & 26.6 & 26.9 & 26.8 & 26.7 & [2] \\
Abell 2667 & 1.5 & 26.5 & 26.9 & 26.8 & 26.7 & [2] \\
Abell 1689 & 5.7 & 28.1 & 26.8 & 26.4 & --- & [3] \\
Abell 1703 & 3.1 & 27.9 & 26.7 & 26.0 & --- & [4] \\
1E0657-56\tablenotemark{e} & 2.7 & 27.6 & 27.2 & 26.0 & --- & [4] \\
Abell 1835 & 0.7 & 27.1 & 27.2 & --- & 26.7 & [4] \\
AC114 & 0.7 & 27.1 & 27.2 & --- & 26.7 & [4] \\
\enddata
\tablenotetext{a}{$5\sigma$ limits assume a $0.3''$-diameter aperture
  for ACS/WFPC2 and $0.6''$-diameter aperture for NICMOS. The source photometry 
  in this paper uses a larger aperture, therefore leading to shallower limiting 
  magnitudes.}
\tablenotetext{b}{The depth of the deepest single-band optical
  ($\leq0.8\mu$m) image available over the cluster.  Note that some of
  the clusters listed here have very deep data in two or more bands
  (particularly for MS1358+61, CL0024+16, Abell 2218, Abell 1689,
  Abell 1703), so the effective depth of the combined optical data is
  often $>$0.4 magnitude deeper than tabulated here.}
\tablenotetext{c}{References: [1] This paper,
[2] \citet{richard}, [3] \citet{tom,zd1}, [4] \citet{bouwens4}.} 
\tablenotetext{d}{These data were previously used by
  \citet{bouwens4} to search for $z$ and $J$ dropout galaxies.  
We use these data to search for bright $i$ dropouts.} 
\tablenotetext{e}{The ``Bullet'' cluster.}
\end{deluxetable}
}
\clearpage 
{
\begin{deluxetable}{lccccccccc}
\rotate
\tablecaption{Photometry of Bright $z \ga 6$ Dropout Objects 
in Cluster Fields (AB Magnitude)\label{tbl-2}}
\tablewidth{0pt}
\footnotesize
\tighttable
\tablehead{
\colhead{Name} & \colhead{$\alpha_{J2000}$} & \colhead{$\delta_{J2000}$} 
& \colhead{$r_{625}$} & \colhead{$i_{775}$} 
& \colhead{$z_{850}$} & \colhead{$J_{110}$}  & \colhead{\H} 
& \colhead{$3.6 \mu$m} & \colhead{$4.5 \mu$m} 
}
\startdata
A1703-iD1\tablenotemark{a} & 13 15 01.41 & +51 48 25.9 &$>28.2$ & $26.8\pm 0.3$& $ 24.2\pm 0.1$& $ 24.0\pm 0.1$ & $23.9 \pm 0.1$ &$23.1 \pm 0.3$&$23.5 \pm 0.4$ \\ 
CL0024-iD1\tablenotemark{a} & 00 26 37.78 & +17 10 40.0  & $> 28.1$ &$ > 27.9$ & $ 26.0\pm 0.2$& $ 25.1\pm 0.1$& $25.0 \pm 0.1$ & $24.4 \pm 0.2$ & $24.4 \pm 0.3$\\ 
CL0024-zD1\tablenotemark{a} & 00 26 37.93 & +17 10 39.0  & $> 28.1$ & $>27.9$ &$ 27.3\pm 0.8$& $ 26.0\pm 0.2$& $ 25.6 \pm 0.2$ & $24.5 \pm 0.5$ & $24.8 \pm 0.5$\\ 
A1689-zD1\tablenotemark{b} & 13 11 29.96 & $-01$ 19 18.7  & $> 27.8$ & $>27.8$ &$ >27.5$& $ 25.3\pm 0.1$& $ 24.7 \pm 0.1$ & $24.2 \pm 0.3$ & $23.9 \pm 0.3$ \\ 
A2218-iD1\tablenotemark{c,d} & 16 35 54.40 & +66 12 32.8  & $> 27.6$ & $>27.2$ &$ 25.1\pm 0.1$& $ 24.3\pm 0.1$& $ 24.1 \pm 0.1$ & $23.7 \pm 0.3$ & $23.9 \pm 0.3$\\ 
\enddata
\tablenotetext{a}{This paper. 
These two sources are only 2.5\arcsec\ apart, have similar redshifts and are
probably spatially associated (See Table 3 and 4 and the text).}
\tablenotetext{b}{\citet{zd1}.}
\tablenotetext{c}{\citet{egami}. The $r$- and $i$-band magnitudes are from our own measurements.}
\tablenotetext{d}{Component b. The other component (a) has a similar 
magnitude, but its IRAC fluxes are unavailable because of contamination from a neighboring source.}
\end{deluxetable}
} 
\clearpage 
\begin{deluxetable}{ccccccc}
\tablecolumns{7}
\tablewidth{0pt}
\tablecaption{Best-fit Stellar-Population Model Results\label{tbl-3}}
\tablehead{ & & &  Mass\tablenotemark{b} & Age\tablenotemark{c} &       &  \\
Object & $z_{\mbox{\scriptsize phot}}$ & $z_{\mbox{\scriptsize form}}$\tablenotemark{a} & ($10^{9}$ $M_{\odot}$) & (Myr)                & $A_V$ & $\chi^{2}_{\nu}$
} 
\startdata
\cutinhead{\sc High Redshift, Solar Metallicity ($Z = 0.02$)} 
A1703-iD1   & $5.9 \pm 0.1$ & 6.2 & $9.2 \pm 3.2$ & $ 55 ^{+17}_{-27}$ & $0.0 \pm 0.2$          & 1.2  \\
A1703-iD1   & 5.827 & 6.2 & $12 \pm 0.6$ & $ 64 ^{+10}_{-9}$ & $0.0 \pm 0.1$          & 1.1  \\
CL0024-iD1  & $6.4 \pm 0.1$ & 6.7 & $1.5 \pm 0.1$ & $ 43 ^{+15}_{-17}$ & $0.0 \pm 0.1$          & 0.7  \\
CL0024-zD1  & $6.5 \pm 0.3$ & 7.1 & $2.0 \pm 0.4$ & $ 81 ^{+41}_{-72}$ & $0.0 \pm 0.1$          & 0.1  \\ 
\cutinhead{\sc High Redshift, Subsolar Metallicity ($Z = 0.0004$)} 
A1703-iD1   & $6.0 \pm 0.1$ & 6.5 & $12 \pm 4.6$  &  $64^{+41}_{-44} $& $0.2 \pm 0.2$          & 1.1  \\
A1703-iD1   & 5.827 & 5.9 & $13 \pm 3.4$  &  $13^{+12}_{-2.5} $& $1.2 \pm 0.1$          & 1.7  \\
CL0024-iD1  & $6.5 \pm 0.1$ & 7.0 & $2.1 \pm 0.8$ &  $72^{+24}_{-44} $& $0.0 \pm 0.2$          & 0.5  \\
CL0024-zD1  & $6.7 \pm 0.4$ & 7.4 & $2.5 \pm 0.2$ &  $90^{+101}_{-77} $& $0.2 \pm 0.4$          & 0.1  \\
\cutinhead{\sc Low Redshift, Solar Metallicity ($Z = 0.02$)} 
A1703-iD1   & $1.2 \pm 0.1$ & 1.4 & $2.4 \pm 0.4$ & $454 ^{+117}_{-49}$ & $0.8 \pm 0.3$          & 9.3  \\
CL0024-iD1  & $1.3 \pm 0.2$ & 1.5 & $0.4 \pm 0.1$ & $571 ^{+235}_{-156}$& $0.3 \pm 0.4$          & 2.8  \\
CL0024-zD1  & $1.4 \pm 0.5$ & 1.5 & $0.4 \pm 0.2$ & $571 ^{+444}_{-315}$& $0.7 \pm 0.9$          & 0.6  \\ 
\cutinhead{\sc Low Redshift, Subsolar Metallicity ($Z = 0.0004$)} 
A1703-iD1   & $1.2 \pm 0.2$ & 1.4 & $4.2 \pm 0.9$ & $454 ^{+140}_{-22} $ & $1.8 \pm 0.3$          & 11.4 \\
CL0024-iD1  & $1.3 \pm 0.2$ & 1.5 & $0.7 \pm 0.2$ & $454 ^{+187}_{-167}$ & $1.4 \pm 0.4$          & 4.2  \\
CL0024-zD1  & $1.4 \pm 0.6$ & 1.6 & $0.6 \pm 0.4$ & $404 ^{+402}_{-290}$& $1.8 \pm 1.0$          & 0.9  \\
\enddata 
\tablenotetext{a}{Formation redshift calculated from the fitted redshift and SSP stellar age.}
\tablenotetext{b}{Best-fit stellar mass, corrected for the cluster magnification.}
\tablenotetext{c}{Age of the single-burst stellar population (SSP model).} 
\end{deluxetable}
\clearpage 
{
\begin{deluxetable}{lccccccc}
\rotate
\tablecaption{Intrinsic Properties of Bright Dropout Objects\label{tbl-4}}
\tablewidth{0pt}
\footnotesize
\tighttable
\tablehead{
\colhead{Name} & \colhead{Photometric} 
& \colhead{Observed Size}& \colhead{Demagnification} 
& \colhead{Luminosity} & \colhead{Size} & \colhead{Component} & \colhead{Relative}
\\
\colhead{} & \colhead{Redshift} & 
\colhead{(\arcsec)} & \colhead{Factor}
& \colhead{($10^{43}$ \ergsec)} & \colhead{(kpc)} & \colhead{Separation (kpc)} & \colhead{Flux}
}
\startdata
A1703-iD1 & $5.95 \pm 0.15$& 0.66 + 0.51\tablenotemark{a} & 3 & $21 $  & 2.2 + 1.7\tablenotemark{a} & 0.9 & 1.0/0.57\\ 
CL0024-iD1 & $6.45 \pm 0.15$ &1.1 + 0.53\tablenotemark{a} & 6.3 & $5.5 $ & 1.0 + 0.5\tablenotemark{a}  & 1.3 & 1.0/0.09\tablenotemark{d}\\ 
CL0024-zD1 & $6.55 \pm 0.35$ & 0.6 + 0.5\tablenotemark{a} & 6.2 & $4.8 $ & 0.5 + 0.4\tablenotemark{a} & 1.4 & 1.0/0.51\tablenotemark{d}\\ 
A1689-zD1 & $7.6 \pm 0.4 $ & 1.6 + 1.0\tablenotemark{a} & 9.3 & 6.9  & 0.9 + 0.5\tablenotemark{a} & 1.6 & 1.0/0.62\tablenotemark{d}\\ 
A2218-iD1\tablenotemark{b} &$6.65 \pm 0.1 $& $3.0 \times 0.3$\tablenotemark{c} & $15 \times 1.7$\tablenotemark{b}  & 3.2  & 0.2 
& \nodata & \nodata \\
\enddata
\tablenotetext{a}{For two separate nuclei.}
\tablenotetext{b}{\citet{egami}. These two sources are only 2.5\arcsec\ apart, 
have similar redshifts and are probably spatially associated (See Table 2 and 
3 and the text).}
\tablenotetext{c}{Along the major and minor axes, as the
image is greatly stretched due to a gravitational lensing effect.}
\tablenotetext{d}{Measured in the $J$ band.}
\end{deluxetable}
}
\clearpage

\begin{thebibliography}{}
\bibitem[Beckwith et al.(2006)]{hudf} Beckwith, S, V. W., et al. 2006, \aj, 132, 1729
\bibitem[Bertin \& Arnouts (1996)]{sex} Bertin E., \& Arnouts S. 1996, \aap, 117, 393
\bibitem[Blakeslee et al.(2003)]{apsis} Blakeslee, J. P., Anderson, K. R., 
Meurer, G. R., Ben\'{\i}tez, N., \& Magee, D. 2003, in Astronomical Data 
Analysis Software and Systems XII (San Francisco: ASP), eds. H. E. Payne, 
R. I. Jedrzejewski, \& R. N. Hook, 295, 257
\bibitem[Bouwens et al.(2006)]{bouwens} Bouwens, R. J., Illingworth, G. D., 
 Blakeslee, J. P., \& Franx, M. 2006, \apj, 653, 53 
\bibitem[Bouwens et al.(2007)]{bouwens3} Bouwens, R. J., Illingworth, G. D., Franx, 
  M., \& Ford, H. 2007, \apj, 670, 928
\bibitem[Bouwens et al.(2003)]{bouwens1} Bouwens, R. J. et al. 2003, \apj, 595, 589
\bibitem[Bouwens et al.(2004$a$)]{bouwensudfp} \underline{\hskip 7em}. 2004$a$, 
\apj, 606, L25 
\bibitem[Bouwens et al.(2004$b$)]{bouwenszd} \underline{\hskip 7em}. 2004$b$, 
  \apj, 616, L79 
\bibitem[Bouwens et al.(2008)]{B08} \underline{\hskip 7em}. 2008, ApJ, 686, 230
\bibitem[Bouwens et al.(2009$a$)]{bouwens4}\underline{\hskip 7em}. 2009$a$, \apj, 
  690, 1764 \bibitem[Bouwens et al.(2009$b$)]{B09}\underline{\hskip 7em}. 2009$b$, in preparation
\bibitem[Bradley et al.(2008)]{zd1}Bradley, L. D. et al. 2008, \apj, 678, 647
\bibitem[Broadhurst et al.(1995)]{br95} Broadhurst, T.~J., 
  Taylor, A.~N., \& Peacock, J.~A.\ 1995, \apj, 438, 49 
\bibitem[Broadhurst et al.(2005)]{tom}Broadhurst, T. et al. 2005, \apj, 621, 53
\bibitem[Bruzual \& Charlot(2003)]{bc03}Bruzual. G. \& Charlot, S. 2003, \mnras, 344, 1000
\bibitem[Bunker et al.(2003)]{bunker} Bunker, A. J., Stanway, E. R., Ellis, R. S., 
McMahon, R. G., \& McCarthy, P. J. 2003, MNRAS, 342, L47
\bibitem[Calzetti {et~al.}(2000)]{Calzetti}{Calzetti}, D., {Armus}, L., {Bohlin}, R.~C., 
{Kinney}, A.~L., {Koornneef}, J., \& {Storchi-Bergmann}, T. 2000, \apj, 533, 682
\bibitem[Chary (2008)]{chary} Chary, R. 2008, \apj, 680, 32
\bibitem[Chen et al.(2009)]{chen} Chen, H.-W., et al.\ 2009, in press,
arXiv:0809.2608 
\bibitem[Dickinson et al.(2004); D04]{goods} Dickinson, M. et al. 2004, \apj, 600, L99
\bibitem[Dow-Hygelund et al.(2005)]{dow} Dow-Hygelund, C. C. et al. 2005, \apj, 630, L137
\bibitem[Egami et al.(2005)]{egami} Egami, E. et al.  2005, \apj, 618, L8
\bibitem[Eyles et al.(2005)]{eyles} Eyles, L., Bunker, A., Stanway, E., Lacy,
  M., Ellis, R., \& Doherty, M. 2005, \mnras, 364, 443 
\bibitem[Frye et al.(2008)]{frye} Frye, B. et al. 2008, \apj, 685, L5
\bibitem[Giavalisco et al.(2004)]{G04} Giavalisco, M. et al. 2004, \apj, 600, L92
\bibitem[Hu et al.(2002)]{hu} Hu, E. M. et al. 2002, \apj, 568, 75
\bibitem[Jee et al.(2007)]{jee} Jee, M. J., et al. 2007, \apj, 661, 728
\bibitem[Kneib et al.(2003)]{kneib3} Kneib, J.-P., et al.\ 2003, \apj, 598, 804
\bibitem[Kneib et al.(2004)]{kneib} Kneib, J.-P., Ellis, R. S., Santos, M. R., \& 
  Richard, J. 2004, \apj, 607, 697
\bibitem[Kron (1980)]{kron} Kron, R. G. 1980, \apjs, 43, 305
\bibitem[Labb\'e et al. (2006)]{labbe} Labb\'e, I,, Bouwens, R. J., 
 Illingworth, G. D., \& Franx, M. 2006, \apj, 649, L67
\bibitem[Limousin et al.(2007)]{limousin7} Limousin, M., et al.\ 
2007, \apj, 668, 643
\bibitem[Limousin et al.(2008)]{limousin}\underline{\hskip 7 em} 2008, \aap, 489, 23 
\bibitem[{{Madau}(1995)}]{madau}{Madau}, P. 1995, \apj, 441, 18
\bibitem[Magee et al.(2007)]{magee}Magee, D. K., Bouwens, R. J. \& Illingworth, G. D. 2007, 
Astronomical Data Analysis Software and Systems XVI, ASP Conference Series, 
eds. R. A. Shaw, F. Hill, \& D. J. Bell, (ASP: San Francisco), 376, 261
\bibitem[Natarajan et al.(2007)]{priya} Natarajan, P., Kneib, 
J.-P., Smail, I., Treu, T., Ellis, R., Moran, S., Limousin, M., 
\& Czoske, O.\ 2007, arXiv:0711.4587 
\bibitem[Oesch et al.(2007)]{oesch} Oesch, P.~A., et al.\ 2007, \apj, 671, 1212 
\bibitem[Ouchi et al.(2008$a$)]{himiko}Ouchi, M. et al. 2008$a$, \apj, submitted (astro-ph/0807.4174)
\bibitem[Ouchi et al.(2008$b$)]{ouchi}\underline{\hskip 7em}. 2008$b$, \apjs, 176, 301
\bibitem[Pell\'o et al.(2004)]{pello}Pell\'o, R., Schaerer, D., Richard, J., 
Le Borgne, J.-F., \& Kneib, J.-P. 2004, \aap, 416, 35
\bibitem[Peng et al.(2002)]{galfit}Peng, C. Y., Ho, L. C., Impley, C. D., \& 
  Rix, H. -W. 2002, \aj, 124, 266
\bibitem[Reddy et al.(2008)]{reddy} Reddy, N.~A., Steidel, C.~C.,
  Pettini, M., Adelberger, K.~L., Shapley, A.~E., Erb, D.~K., \&
  Dickinson, M.\ 2008, \apjs, 175, 48
\bibitem[Richard et al.(2006)]{richard06}Richard, J. et al. 2006, \aap. 456, 861
\bibitem[Richard et al.(2008)]{richard}\underline{\hskip 7em}. 2008, \apj, 685, 705
\bibitem[Richard et al.(2009)]{a1703}\underline{\hskip 7em}. 2009, astro-ph/0901.0427
\bibitem[{Salpeter(1955)}]{imf}Salpeter, E.~E. 1955, \apj, 121, 161
\bibitem[Schaerer \& Pell\'o(2005)]{schaerer}Schaerer, D. \& Pell\'o, R. 2005, \mnras, 362, 1054
\bibitem[Shimasaku et al.(2005)]{subaru} Shimasaku, K., Ouchi, M., Furusawa, 
  H. Yoshida, M., Kashikawa, N., \& Okamura, S. 2005, PASJ, 57, 447 
\bibitem[Stark et al.(2007)]{stark} Stark, D. P., Bunker, A. J., Ellis, R. S., Eyles, L. P., 
  \& Lacy, M. 2007, \apj, 659, 84
\bibitem[Stanway et al.(2004)]{stanway1}Stanway, E., Bunker, A., McMahon, R., Ellis, R. S., 
 Treu, T. \& McCarthy, P. 2004, \apj, 607, 704
\bibitem[Stanway et al.(2005)]{stanway2}Stanway, E. R., McMahon, R. G., \& 
 Bunker, A. J. 2005, \mnras, 359, 1184
\bibitem[Szalay et al.(1999)]{szalay} Szalay, A.~S.,
Connolly, A.~J., \& Szokoly, G.~P.\ 1999, \aj, 117, 68
\bibitem[Taniguchi et al.(2005)]{taniguchi} Taniguchi, Y., et al. 2005, PASJ, 57, 165
\bibitem[Thompson et al.(2005)]{thompson} Thompson, R. I., et al. 2005, AJ, 130, 1 
\bibitem[Yan \& Windhorst(2004)]{yan4} Yan, H.~\& Windhorst, R.~A.\ 2004, \apj, 612, L93
\bibitem[Yan et al.(2005)]{yan5} Yan, H., et al.\ 2005, \apj, 634, 109
\bibitem[Yan et al.(2006)]{yan} Yan, H., Dickinson, M., Giavalisco, M.,
Stern, D., Eisenhardt, P. R. M., \&  Ferguson, H. C. 2006, \apj, 651, 24
\bibitem[Yoshida et al. (2006)]{yoshida} Yoshida, M., et al.\ 2006, \apj, 653, 988
\bibitem[Zheng et al.(2009)]{zheng} Zheng, W. et al. 2009, in preparation
\bibitem[Zitrin et al.(2009)]{adi} Zitrin, A. et al. 2009, arXiv:0902.3971
\end{thebibliography}
\end{document}